\documentclass[useAMS,usenatbib]{mn2e}
\usepackage{graphicx, rotating}
\usepackage{aas_macros}  
\usepackage{graphicx}
\usepackage{amssymb}
\usepackage{rotating}
\usepackage{multirow}
\usepackage[breaklinks,colorlinks=true,citecolor=black,linkcolor=black,urlcolor=blue,bookmarks=true]{hyperref}

\usepackage{ulem}

\newcommand{\kms}{km\,s$^{-1}$}
\newcommand{\HI}{\textrm{H}\,\textsc{i}}
\newcommand{\HII}{\textrm{H}\,\textsc{ii}}

\title[A new isolated dSph galaxy near the Local Group]{A new isolated dSph galaxy near the Local Group
}
\author[I. D. Karachentsev et al.]{
I. D. Karachentsev$^{1}$\thanks{E-mail: ikar@sao.ru}, 
L. N. Makarova$^{1}$,
D. I. Makarov$^{1}$,
R. B. Tully$^{2}$,
L. Rizzi$^3$\\
$^{1}$Special Astrophysical Observatory, Nizhniy Arkhyz, Karachai-Cherkessia 369167, Russia\\
$^{2}$Institute for Astronomy, University of Hawaii, 2680 Woodlawn Drive, HI 96822, USA\\
$^{3}$W. M. Keck Observatory, 65-1120 Mamalahoa Hwy, Kamuela, HI 96743, USA
}
\begin{document}

\date{Accepted XXX. Received XXX; in original form XXX}

\pagerange{\pageref{firstpage}--\pageref{lastpage}} \pubyear{XXX}

\maketitle

\begin{abstract}
Observations of the highly isolated dwarf spheroidal galaxy KKs\,3 = [KK2000]\,03
with the Hubble Space Telescope (HST) Advanced Camera for Surveys (ACS)
are presented. We measured the galaxy distance of $2.12\pm0.07$ Mpc using 
the Tip of Red Giant Branch (TRGB) method. 
The total blue absolute magnitude of the galaxy is estimated as $M_B = -10.8$ mag.
We briefly discuss the star formation history of KKs\,3 derived from its colour-magnitude diagram.
According to our calculation, the total stellar mass of the galaxy is $2.3\times 10^7M_{\odot}$, and
most stars (74\%) were formed at an early epoch more than 12 Gyr ago.
A full description of the properties of the colour-magnitude diagram requires some extension of star formation in metallicity and age.
\end{abstract}


\section{Introduction} 

\label{firstpage}

  In the standard cosmological model, $\Lambda$CDM, the formation of galaxies
begins with the hierarchical merging of dwarf objects \citep{white1978}.
The dwarf galaxies that survive until today are sensitive to
their environment, given their shallow potential wells and characteristic
internal motions of 5 to 30 \kms.
Gas-rich dwarf irregular (dIrr) galaxies with
active star formation are common in the
general field and the outer regions of groups.
The gas-poor dwarf spheroidal systems (dSph) with old stellar
population are found almost exclusively in the virial domain of groups and clusters.
It is generally considered that the concentration of dSph galaxies to
the halos of massive galaxies is due to processes of gas stripping and strangulation
of irregular dwarfs that limits further star formation, transforming dIrr
into dSph. If these mechanisms that are operative in dense environments 
are paramount then
dSph objects should be absent in the general field of low densities.
However, energetic events associated with
active star formation in dwarf systems
at an early epoch might deplete gas resources. In such
cases, relic quenched dSph galaxies may occur among isolated objects.
Alternatively, \citet{benitez2013} offer an
intermediate mechanism, with the transformation of dIrr galaxies into dSphs
occurring any time via so-called ``cosmic web stripping''.

   Beyond the issue of the relationship between dIrr and dSph,
the search and discovery of isolated spheroidal dwarfs constitutes
an important interest for cosmology, given the unexpectedly small number of 
dwarfs of any type. Over the last decade, about 30 dSph galaxies have been 
discovered in the Local Group through systematic
search in the vicinity of M\,31 \citep{PAndAS1} and
more that a dozen dSphs were discovered by \citet{chiboucas2009}
in the nearby group around M\,81.  The discoveries resulted from targeted searches
within small parts of the sky ($\sim 390$ deg$^2$ and $\sim 65$ deg$^2$, 
respectively). The hunt for isolated spheroidal dwarfs is very difficult 
because it requires a survey of large sky area and considerable sensitivity. 
Objects devoid of neutral hydrogen and high contrast \HII-regions are usually
invisible in optical and \HI-surveys. Only very nearby dSph galaxies, inevitably
of low surface brightness, may be revealed if they resolve into
individual stars.

   Since 2008, only three galaxies had been newly discovered in a spherical 
shell between radii 1 and 3 Mpc around the Local Group. Two of them are dIrrs, 
UGC\,4879 \citep{kopylov2008} and Leo P \citep{giovanelli2013}, and the third 
one, KK\,258 \citep{karachentsevKK258}, belongs to the transition type dTr with 
minimal but detectable gas and young stars. Here we report the discovery in 
this volume of a dwarf spheroidal system KKs\,3 ([KK2000]\,03 = SGC\,0224.3--7345 
in the nomenclature of the NASA/IPAC Extragalactic Database) at a distance of 
$D = 2.12\pm0.07$ Mpc and well removed from any other known galaxy.

\section{ACS HST observations and TRGB distance}

\begin{figure}
\includegraphics[width=9cm]{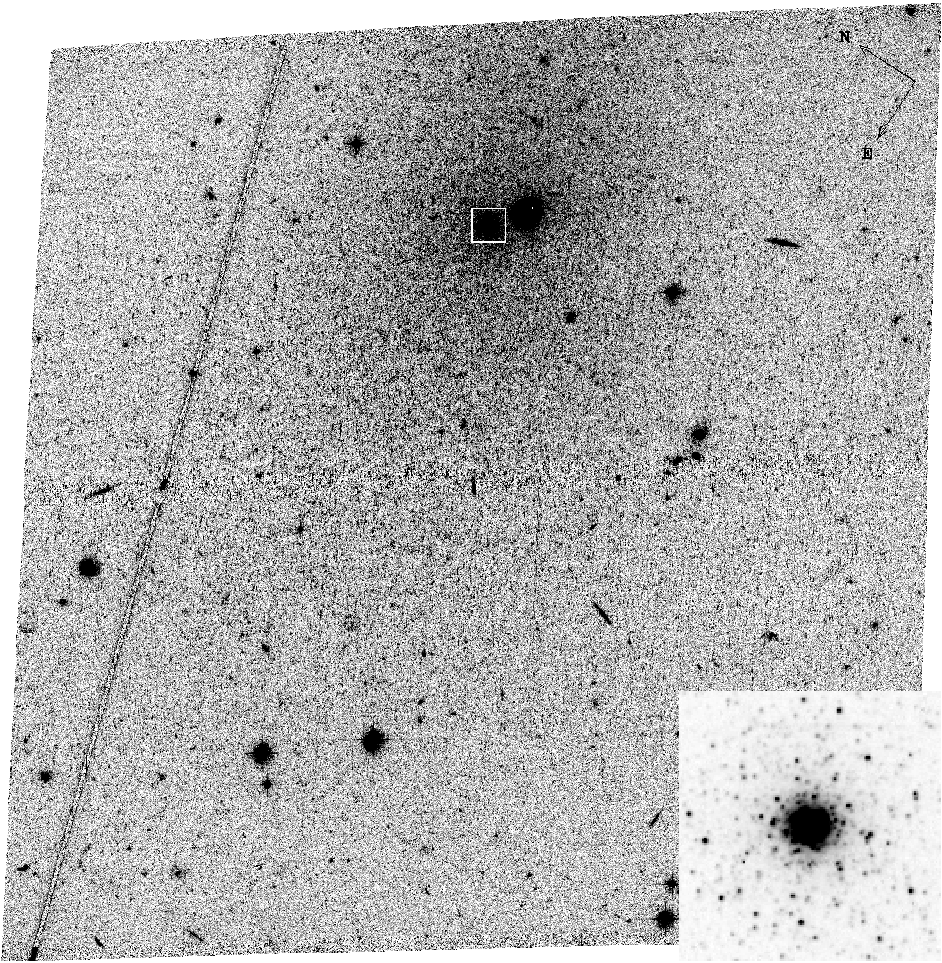}
\caption{\textit{HST}/ACS image of KKs\,3 through the \textit{F606W} filter. 
The image size is $3.4\times3.4$ arcmin. 
The 7.5 arcsec region highlighted by the white square near the centre of the 
galaxy is shown in the lower right corner to contain a globular cluster. 
}
\label{fig:ima}
\end{figure}

The low surface brightness object KKs\,3 with J2000 coordinates: 
R.A.\ = $02^h24^m44\fs4$, Dec.\ = $-73^{\circ}30\arcmin51\arcsec$ was 
detected in full sky surveys by \citet{kk2000} and \citet{whiting2002}
as a potential dSph galaxy neighbouring the Local Group. Even earlier,
the object was recorded by \citet{corwin1985}.
In HyperLeda\footnote{\url{http://leda.univ-lyon1.fr}} the galaxy is listed as 
PGC\,09140. The `Updated Nearby Galaxy Catalog' \citep{karachentsev2013} 
assigns KKs\,3 the total apparent magnitude  $B = 16.0$ mag and the Holmberg 
diameters $2.5\times 1.0$ arcmin. The assigned distance of 4 Mpc in that 
catalogue erroneously assumed an association 
of KKs\,3 with the galaxy NGC\,1313. \citet{kirby2008} imaged KKs\,3 at the 
3.9-m AAT telescope in H-band and noted it as almost invisible with an 1800 s 
exposure.

The dwarf  galaxy KKs\,3 was observed aboard \textit{HST} using Advanced Camera
for Surveys (ACS) on 29 August, 2014 (SNAP 13442, PI R. Tully). Two exposures 
were made in a single orbit with the filters \textit{F606W} (1200 s) and 
\textit{F814W} (1200 s). The \textit{F606W} image of the galaxy  KKs\,3 is 
shown in Fig.~\ref{fig:ima}. 
In the central part of the galaxy, highlighted 
by the white square, we found a globular cluster shown in the lower right corner
of the Fig.~\ref{fig:ima}. Photometry of the cluster yields its total magnitude $V = 18.30\pm0.03$
mag and the colour $V-I = 0.89\pm0.06$. The bright spot at the center of the
galaxy, just to the right of the globular cluster box is caused by a tight pair
of red foreground stars.

\begin{figure}
\includegraphics[height=10cm]{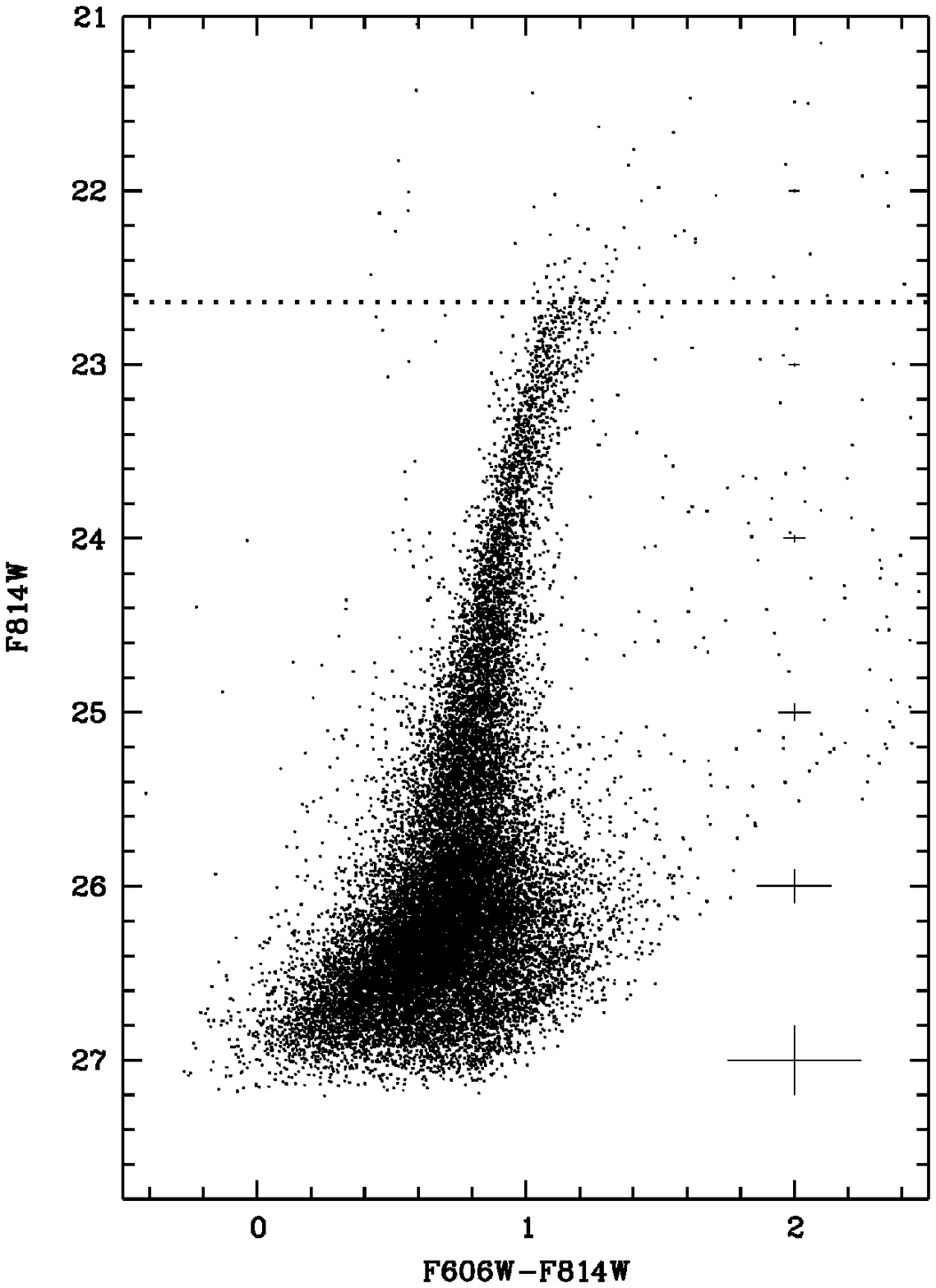}
\caption{Colour-magnitude diagram of KKs\,3. 
Photometric errors are indicated by the bars at the right side 
of the CMD. The TRGB position is indicated by the dotted line.
}
\label{fig:cmd}
\end{figure}

The photometry of resolved stars in the galaxy was performed with the ACS module
of the \textsc{DOLPHOT} package\footnote{http://purcell.as.arizona.edu/dolphot/}
for crowded field photometry \citep{dolphin02} using the recommended recipe and
parameters. Only stars with photometry of good quality were included in
the final compilation, following recommendations given in the \textsc{DOLPHOT} 
User's Guide.

Artificial stars were inserted and recovered using the same reduction procedures
to accurately estimate photometric errors, including crowding and blending effects.
A large library of artificial stars was generated spanning
the full range of observed stellar magnitudes and colours to assure that the 
distribution of the recovered photometry is adequately sampled.

We have determined the KKs\,3 distance with
our \textsc{trgbtool} program which uses a maximum-likelihood algorithm
to determine the magnitude of the tip of the red giant branch (TRGB) from the 
stellar luminosity function \citep{makarov2006}. The estimated value of the TRGB 
is $\textit{F814W} = 22.64\pm0.04$ mag in the ACS instrumental system.
Following the calibration of the TRGB methodology developed by \citet{rizzi2007},
we have obtained the true distance modulus $(m-M)_0 = 26.63\pm0.07$ mag and the 
distance of $D = 2.12\pm0.07$ Mpc. This measurement assumes foreground reddening 
of E(B-V) = 0.045 \citep{sf2011}.

\section{Colour-magnitude diagram and star formation history}

\begin{figure*}
\includegraphics[height=7cm]{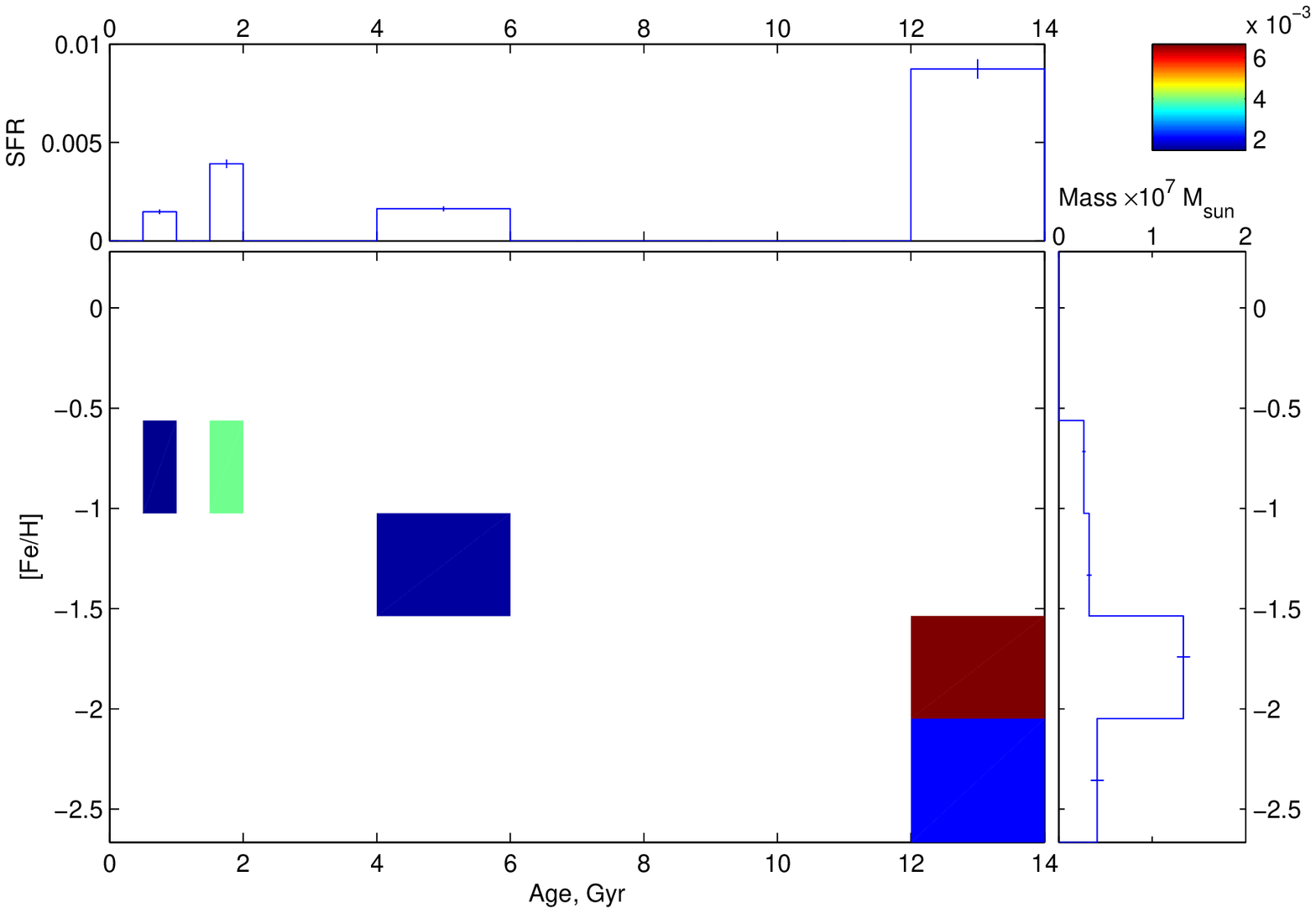}
\hspace{3mm}
\includegraphics[height=7cm]{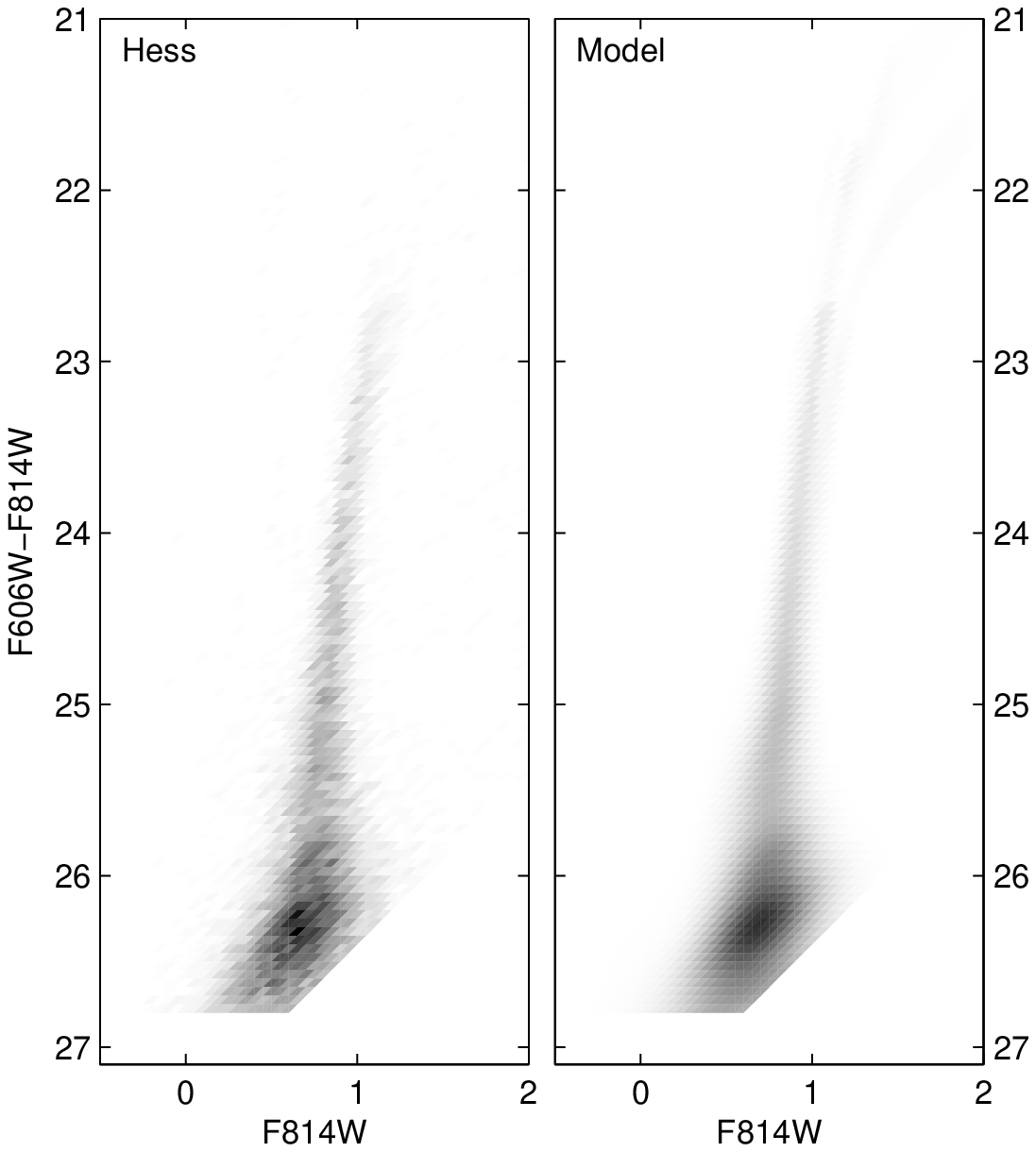}
\caption{
Left plot:
The star formation history of KKs\,3.
The top panel shows the star formation rate (SFR) ($M_\odot$\,yr$^{-1}$) 
against the age of the stellar populations.
The bottom panel shows the metallicity of the stellar components as a function of age. 
The side panel plots stellar mass vs.\ metallicity.
The resulting errors in the SFR are indicated with the vertical bars.
These errors only represent the statistical uncertainties of the fit.
Right plot:
Observational (left) and reconstructed (right) colour-magnitude diagrams of KKs\,3.
The grey-scale density encodes the number of stars in a bin.
}
\label{fig:sfh}
\end{figure*}

Figure~\ref{fig:cmd} shows the colour-magnitude diagram (CMD) of KKs\,3. 
The measured TRGB position is marked by the dotted line. The variety of 
resolved stellar populations is not large in this picture. We can see a tight 
and well populated red giant branch (RGB) and a clearly visible red clump 
toward the bottom of the RGB in the magnitude range 
$25.8 \leq \textit{F814W} \leq 26.8$. Asymptotic giant branch (AGB) stars
in the one magnitude interval brighter than the TRGB are scarce. The few stars 
with the colour index of about 0.5--0.6 and \textit{F814W} between 21 and 24 
mag are most likely foreground objects. They are randomly scattered 
on the ACS image rather than concentrate on the galaxy body. 
The lack of manifestations of star formation within the last Gyr puts an 
extreme limit on recent activity. According to the colour-magnitude diagram, 
the galaxy under study should be classified as a typical dwarf spheroidal.
\HI{} is undetected although the line-of-sight is confused by the Magellanic Stream.

We determine the quantitative star formation and metal enrichment history 
of KKs\,3 from the CMD using our \textsc{StarProbe} program. The program 
develops an approximation to the observed distribution of stars in the CMD 
using a positive linear combination of synthetic diagrams formed by simple 
stellar populations (SSP: sets of single age and single metallicity 
populations). We used the Padova2000 theoretical isochrones \citep{girardi00} 
and a \citet{salpeter55} initial mass function (IMF). 
The synthetic diagrams were altered by the same
incompleteness, crowding effects, and photometric systematics
as those determined for the observations using artificial star
experiments. A full description of the details of our approach and 
the {\sc StarProbe} software are given in the papers of \citet{mm4, makarova2010}.
Figure~\ref{fig:sfh} demonstrates the result of our calculations of 
the star formation history (SFH) of the KKs\,3 dwarf galaxy.
The uncertainty of each star formation episode is estimated from an analysis of 
the likelihood function. Upper and lower error bars are calculated for a 
significance level of 0.317, which corresponds to $1\sigma$ for a normal 
distribution. The errors only reflect statistical uncertainties of the fit. 
Systematic uncertainties caused by the isochrone set, the IMF choice, 
the distance and reddening estimates are not included in error bars.

Three star formation episodes are distinguished.
The most prominent old star burst occurred 12--14 Gyr ago, 
the middle-age population was formed about 5 Gyr ago 
and there is evidence for the existence of even
younger stars with age $\leq2$ Gyr.
Note that the model anticipates more AGB stars than are seen in the observed CMD.
The clearly visible red clump, helium burning stars at the base of the AGB,  
plays an important role in the age and metallicity reconstruction.
The oldest population ($\sim13$ Gyr) gives a red clump that is too faint. 
The presence of a younger component populates a brighter red clump in 
accordance with the observations. However, the contribution of this younger 
population is small in stellar mass and star formation rate. A lower limit 
in age is established by the absence of main sequence or blue loop features.

According to our calculations, the total stellar mass of KKs\,3
is $2.3\times$10$^7\,M_{\odot}$, and most of the stars were formed
in the early epoch of the Universe 12--14 Gyr ago. We estimate the mass 
fraction formed during this initial event to be 74 per cent.
The average rate of star formation in this period was high, 
$8.7\pm0.4\times10^{-3}$\,$M_{\odot}$\,yr$^{-1}$. 
It can be seen from the Fig.~\ref{fig:sfh}, that the average metallicity 
of the primordial stars is extremely low, [Fe/H]$\simeq-1.9$.
Less intensive star formation is noticeable about 4--6 and 0.8--2 Gyr ago. These
episodes contribute about 14 per cent and 12 per cent of the total stellar mass, 
respectively. There is metal enrichment of the stars with time, with
the youngest stars, at about 1 Gyr, most likely
metal-enriched to [Fe/H]$\simeq-0.7$.

\section{Environment of KKs\,3 and other nearby dSphs}

\begin{figure*}
\includegraphics[width=18cm]{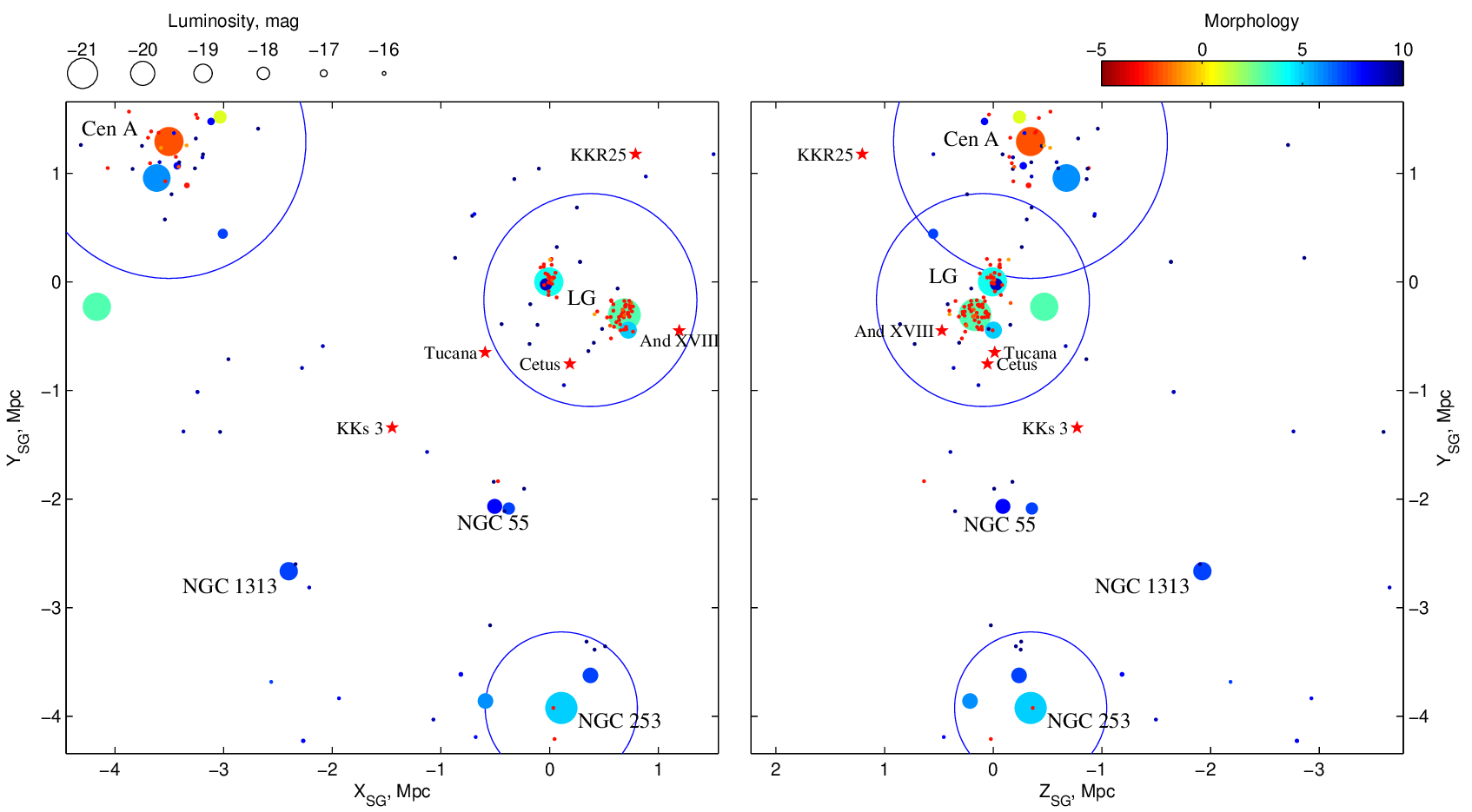}
\caption{Landscape of neighbouring galaxies around KKs\,3.
The colour and size of the circles represent the morphological type and 
luminosity of a galaxy according to the given scales.
Three circles outline spheres of zero-velocity surfaces around the
three massive groups: the Local Group and groups around NGC\,5128 (Cen\,A) 
and NGC\,253. Red stars locate five dSph galaxies that in varying degrees 
are isolated systems.
}
\label{fig:env}
\end{figure*}

The distribution of neighbouring galaxies in a cubic volume $\pm3$ Mpc 
around KKs\,3 is presented in the Fig.~\ref{fig:env} in two Cartesian 
Supergalactic projections. The colour and size of the circles represent 
the morphological type and 
luminosity of the galaxies according to the given scales. In this volume
there are three massive groups: the Local Group and groups around NGC\,5128
(Cen\,A) and NGC\,253. Three circles outline spheres of zero-velocity
surfaces. Inside these domains group members do not participate in the cosmic
expansion. At a little over 1 Mpc from KKs\,3 there are also two small groups,
one around NGC\,55 and the other around NGC\,1313. The nearest neighbours to 
our object are the following dwarf galaxies: IC\,3104 ($M_B=-14.8$ mag, at 
0.99 Mpc), ESO\,294--010 ($M_B=-10.2$ 
mag, at 1.21 Mpc), IC\,5152 ($M_B=-15.6$ mag, at 1.23 Mpc) and Tucana ($M_B=-9.2$ mag, 
at 1.34 Mpc).
 Red stars in Fig.~\ref{fig:env} locate five dwarf spheroidal galaxies
that in varying degrees are isolated systems. Two of them, Cetus and 
And\,XVIII, are located inside the zero-velocity sphere of the Local Group, 
and a third, Tucana, lies just outside this surface. Only two dSph galaxies 
in the volume under consideration are cleanly isolated objects: KKR\,25 at a 
distance of $D_{MW}= 1.93$ Mpc \citep{kar2001,mak2012} and KKs\,3.
One can affirm with virtual certainty that neither of these spheroidal 
dwarfs have experienced a disturbance from their neighbours over the last 
$\sim10$ Gyr.

  Apart from the two true isolated spheroidal dwarfs in the vicinity of 
our Local Group, the local volume within $D < 10$ Mpc contains one more 
known isolated dSph object, Apples\,I. That galaxy was found accidentally 
by \citet{pasq2005} on an image obtained with ACS HST. The radial velocity 
of Apples\,I with respect to the Local Group frame is 661 \kms{} and its 
distance of $8.3\pm2.2$ Mpc is derived from spectroscopic classification 
of the brightest stars. Since the detection of such objects is difficult, 
the number of them within 10 Mpc may be considerable.

  Recently there have been discussions of two rather deep multi-band surveys 
over areas of 300 square degrees \citep{jiang2014} and 170 square degrees
\citep{hil2014}. The limiting magnitudes of these surveys could
resolve the brightest RGB stars of dSph galaxies if they lie within 4 and 
5 Mpc, respectively. Inspection of these images fails to reveal any
isolated dSph candidates. Given the relative sky fraction of these surveys
($\sim1$ per cent) and their depth, the expected total number of isolated 
spheroidal dwarfs in the local volume does not exceed  $N\sim900$. However, 
this quantity corresponds closely with the total number of known galaxies 
in the local volume. Therefore, the present-day observational data do not 
reject the presence of a significant population of relic ``quenched'' dwarfs 
comparable in number with all other kinds of galaxies. Future deep wide-field 
surveys of the sky, like Pan-STARRS1 \citep{tonry2012}, are required to bring 
clarity to this important aspect of the problem of galaxy formation and 
evolution.

\section{Summary}

The key results can be enumerated.
(1) The dominance of ancient stars and the lack of any evidence of recent 
star formation in the CMD justifies the classification of KKs\,3 as a dSph.
(2) KKs\,3 is 2 Mpc from the nearest large galaxy and 1 Mpc from any 
known dwarf so qualifies as isolated.
(3) Although a very old and metal poor population is dominant, 
this early episode of star formation is probably not enough to explain 
the breadth of the RGB and the luminosity range of the red clump.
(4) KKs\,3 sustained limited star formation over many Gyr after its early 
development but exhausted its star forming fuel in isolation. 
 
 \section*{Acknowledgements}
The authors thank Nicolas Martin for constructive comments which helped improve the article.
This work is based on observations made with the NASA/ESA Hubble Space Telescope.
STScI is operated by the Association of Universities for Research in Astronomy, Inc. under NASA contract NAS 5--26555.
The work in Russia is supported by RFBR grants 13--02--90407 and 13--02--92960.
We acknowledge the support from RFBR grant 13--02--00780 and Research 
Program OFN--17 of the Division of Physics, Russian Academy of Sciences.

\bibliographystyle{mn2e}
\bibliography{kks03}   

\bsp
\label{lastpage}

\end{document}